\documentclass{article}
\usepackage{spconf,amsmath,graphicx}
\usepackage{caption}
\usepackage{booktabs}
\usepackage{multirow}
\usepackage{adjustbox}
\usepackage{hyperref}
\usepackage{marvosym}
\usepackage{algorithm}
\usepackage{amssymb}
\usepackage{algorithmicx}
\usepackage{threeparttable}
\usepackage[bottom]{footmisc}

\title{neural speaker diarization using memory-aware multi-speaker embedding with sequence-to-sequence architecture}
\name{
    Gaobin Yang$^1$,
    Maokui He$^1$,
    Shutong Niu$^1$,
    Ruoyu Wang$^1$,
     Yanyan Yue$^2$,
    Shuangqing Qian$^2$,
}

\secondlinename{
    Shilong Wu$^1$,
    Jun Du$^1$,
    Chin-Hui Lee$^3$
}

\address{$^1$University of Science and Technology of China, Hefei, China \\
         $^2$iFlytek Research, Hefei, China \\
         $^3$Georgia Institute of Technology, Atlanta, GA, USA 
}
\begin{document}
%
\maketitle
%
\begin{abstract}
We propose a novel neural speaker diarization system using memory-aware multi-speaker embedding with sequence-to-sequence architecture (NSD-MS2S), which integrates the strengths of memory-aware multi-speaker embedding (MA-MSE) and sequence-to-sequence (Seq2Seq) architecture, leading to improvement in both efficiency and performance. Next, we further decrease the memory occupation of decoding by incorporating input features fusion and then employ a multi-head attention mechanism to capture features at different levels. NSD-MS2S achieved a macro diarization error rate (DER) of 15.9$\%$ on the CHiME-7 EVAL set, which signifies a relative improvement of 49$\%$ over the official baseline system, and is the key technique for us to achieve the best performance for the main track of CHiME-7 DASR Challenge. Additionally, we introduce a deep interactive module (DIM) in MA-MSE module to better retrieve a cleaner and more discriminative multi-speaker embedding, enabling the current model to outperform the system we used in the CHiME-7 DASR Challenge. Our code will be available at \url{https://github.com/liyunlongaaa/NSD-MS2S}.
\end{abstract}
\section{Introduction}
\label{sec:intro}
Speaker diarization is the task of tagging an audio recording with labels that indicate ``who spoke when'' \cite{review1}. High-quality diarization outcomes can be beneficial for numerous speech-related tasks, such as generating meeting summaries, analyzing phone conversations, transcribing dialogues, and so on \cite{review2}. However, speaker diarization remains challenging in real-world scenarios with varying speaker numbers, adverse acoustic environments, and a large portion of speech overlap.

Conventional clustering-based methods \cite{CD1}, which include voice activity detection (SAD/VAD), speech segmentation, speaker feature extraction (e.g., i-vector \cite{ivector}, d-vector \cite{dvector}, x-vector \cite{xvector}), speaker clustering \cite{AHC, COSINE, SC} and re-segmentation \cite{VBResegmentation1}, are commonly used in speaker diarization task. Although clustering-based speaker diarization is relatively robust across different domains, they cannot deal well with overlap segments because every segment can only be assigned a single label through clustering.


 To tackle this problem above, many excellent methods \cite{EEND2, PRN, TSVAD, MAMSE, seq2seq} have been proposed.  Recently, target-speaker voice activity detection (TS-VAD) \cite{TSVAD}\cite{he2021targetspeaker} was proposed, which used speech features along with speaker embedding as input to predict each speaker’s activities at each frame. Although TS-VAD has been a great success in many scenarios \cite{Dihard-III}, it still has some unresolved problems. First, the BLSTM-based network architecture leads to slower inference and significantly increases GPU memory usage as decoding duration becomes longer.  Second, TS-VAD uses a pre-trained extractor to obtain speaker embeddings (e.g. i-vectors) as input, but the embeddings are often unreliable in real scenarios due to the lack of oracle speaker segments. To solve the computational overhead problem mentioned above, Seq2Seq-TSVAD \cite{seq2seq} introduced the sequence-to-sequence architecture and achieved good results. Our previous work NSD-MA-MSE \cite{MAMSE} was proposed to solve the problem of unreliable speaker embeddings in real scenarios, which introduced a memory module to facilitate a dynamical refinement of speaker embedding to reduce a potential data mismatch between the speaker embedding extraction and the NSD network. However, the first problem mentioned above remains unsolved in NSD-MA-MSE. 

In order to further address both of the TS-VAD problems mentioned above simultaneously, in this paper, we propose a novel neural speaker diarization system for CHiME-7 referred to NSD-MS2S, which first integrates the strengths of MA-MSE and Seq2Seq architecture perfectly. NSD-MS2S processes acoustic features and multi-speaker embeddings separately to avoid dimensional expansion. It then combines these two components through a decoder to predict voice activities for the target speakers, resulting in a significant improvement in both efficiency and performance compared to NSD-MA-MSE \cite{MAMSE}. Then, we introduce a simple and efficient method of input features fusion to further reduce the computational overhead required in the decoder of the transformer and then use a multi-head attention mechanism to capture features at different levels. NSD-MS2S achieved a macro DER of 15.9$\%$ on the CHiME-7 EVAL set, representing a relative improvement of 49$\%$ over the official baseline system and enabling us to secure first place in the main track of the CHiME-7 DASR Challenge.  Additionally, in order to better retrieve multi-speaker embedding from the memory module, we introduce a deep interactive module (DIM) in MA-MSE module. Multi-scale feature fusion of acoustic features and speaker embedding basis vectors can retrieve a cleaner and more discriminative multi-speaker embedding than the original MA-MSE module, making the single model results outperform the system we used in the CHiME-7 DASR Challenge.  


\section{METHODS}
\label{sec:methods}


\begin{figure*}[htbp]
    \centering
    

    \includegraphics[width=6.2in]{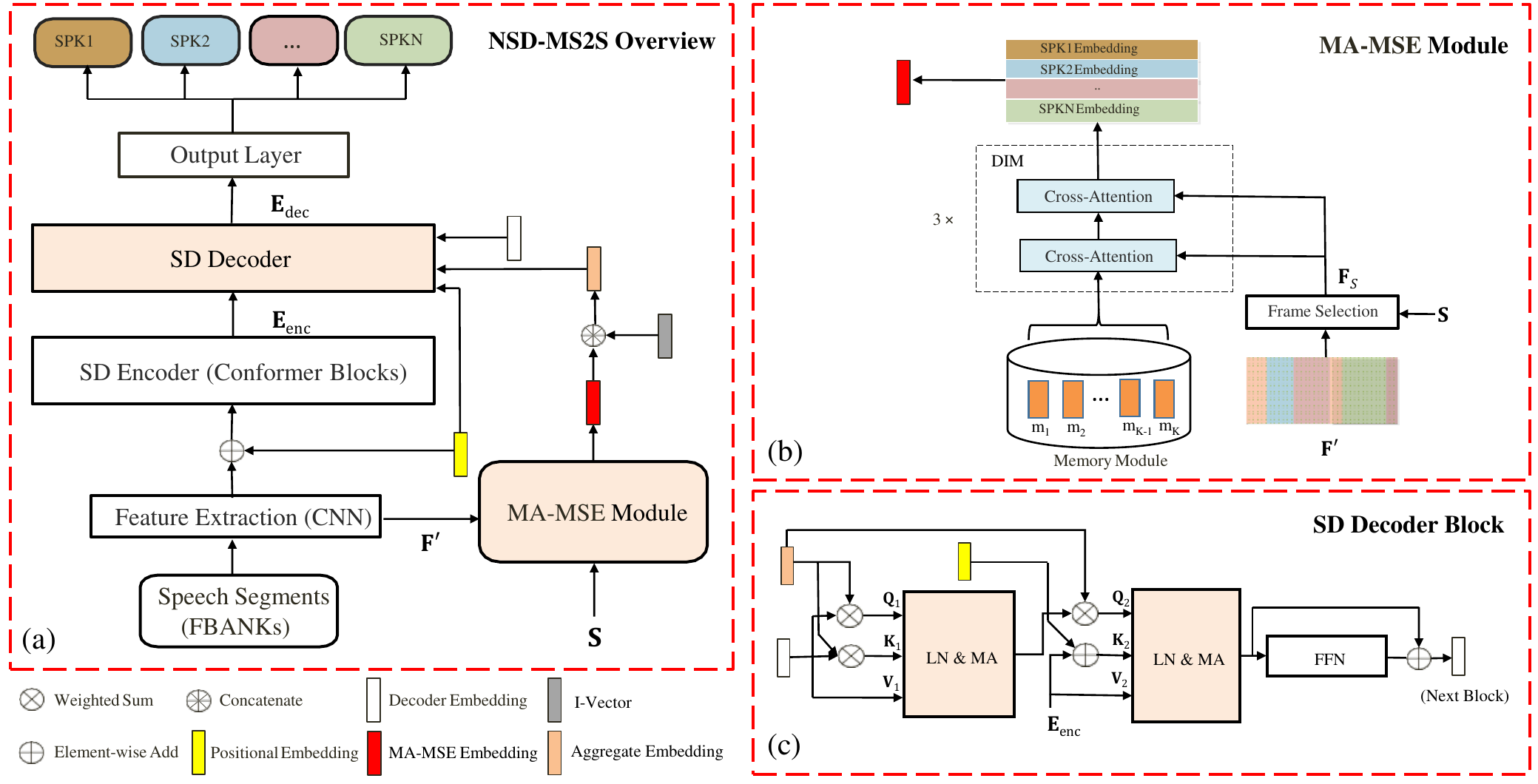}
    \caption{The proposed NSD-MS2S framework. (a) is the overview of NSD-MS2S architecture. (b) and (c) are schematic diagrams of the structures of MA-MSE Module and SD Decoder Block, respectively. }
    \label{fig:NSD-MS2S}
\end{figure*}
The framework of our proposed NSD-MS2S system is shown in Fig.\ref{fig:NSD-MS2S}. The details are introduced as follows.

\vspace{-0.2cm}
\subsection{Overview of Network}
The input of the main network is a set of acoustic features denoted as $\mathbf{{X}} \in \mathbb{R}^{T \times F}$, where $T$ and $F$ are the length and dimension of log-Mel filter-bank features (FBANKs), respectively. Then, the convolutional layers are used to extract a set of deep features denoted by $\mathbf{F} \in \mathbb{R}^{C \times T \times \frac{F}{2}}$, which is downsampled to $\mathbf{F'} \in \mathbb{R}^{ T \times D}$, where $C$ and $D$ are dimensions of channels and features, respectively. The features sequence $\mathbf{F'}$ with positional embedding (PE) is encoded into $\mathbf{E}_{\text{enc}} \in \mathbb{R}^{ T \times D}$ by the following speaker detection (SD) encoder, which is a stack of conformer blocks. In addition, $\mathbf{F'}$ and speaker mask matrix $\mathbf{S} \in \mathbb{R}^{N \times T}$  serve as inputs of MA-MSE module, and then the output MA-MSE embedding $\mathbf{E}_M \in \mathbb{R}^{N \times D_M}$ is obtained, where $N$ is the number of speakers and $D_M$ is the dimension of MA-MSE embedding. We concatenate MA-MSE embedding and i-vector to get aggregate embedding $\mathbf{E}_A \in \mathbb{R}^{N \times D}$. We will give specific details about $\mathbf{E}_A$ in Section \ref{IDM}.Then, we pass aggregate embedding $\mathbf{E}_A$, decoder embedding $\mathbf{E}_D \in \mathbb{R}^{N \times D}$ and $\mathbf{E}_{\text{enc}}$ along with sinusoidal positional embedding to SD decoder to produce $\mathbf{E}_{\text{dec}} \in \mathbb{R}^{ N \times D}$. We will provide a more detailed description of this process in Section \ref{SD_Decoder}. Finally, the output layer can transform $\mathbf{E}_{\text{dec}}$ into posterior probabilities $\mathbf{\hat{Y}} = \left [ \hat{y}_\text{nt} \right ]_{N \times T}$ of voice activities for $N$ speakers. 

\vspace{-0.3cm}
\subsection{Speaker Detection Decoder}
\label{SD_Decoder}
The design of the speaker detection (SD) decoder was mainly inspired by \cite{seq2seq, dert}. The SD decoder consisting of several SD blocks predicts target-speaker voice activities by considering cross-speaker correlations. 

For the forward process of a SD block, first, decoder embedding $\mathbf{E}_D$ and aggregate embedding $\mathbf{E}_A$ pass through their respective multi-layer perception (MLP) to generate within-block representations $\mathbf{E}_D^{Q_1}$, $\mathbf{E}_D^{K_1}$, $\mathbf{E}_D^{V_1}$, $\mathbf{E}_A^{Q_1}$ and $\mathbf{E}_A^{K_1}$. $Q$, $K$, and $V$ denote the query, key, and value in the attention mechanism, respectively. If not specifically noted, all the MLP layers map dimension sizes of input features to $D$. For simplicity, the MLP structure is omitted in Fig.\ref{fig:NSD-MS2S}(c). Then, in order to make the decoder embedding contain speaker information and reduce the subsequent time and space overhead, we fuse the input features and do not increase the feature dimensions, which can be represented by the following equation:
\begin{equation}
    \mathbf{Q}_1 = \mathbf{\beta}_1 \times \mathbf{E}_D^{Q_1} + \left ( 1 - \mathbf{\beta}_1 \right ) \times \mathbf{E}_A^{Q_1} \nonumber
\end{equation}
\begin{equation}
    \mathbf{K}_1 = \mathbf{\beta}_2 \times \mathbf{E}_D^{K_1} + \left ( 1 - \mathbf{\beta}_2 \right ) \times \mathbf{E}_A^{K_1}
\end{equation}
\begin{equation}
    \mathbf{V}_1 = \mathbf{E}_D^{V_1} \nonumber
\end{equation}
where $\mathbf{\beta}_1$ and $\mathbf{\beta}_2$ are learnable parameters, we expect the model itself to decide what information it needs. $\mathbf{Q}_1$, $\mathbf{K}_1$, and $\mathbf{V}_1$ sequentially go through layernorm (LN) and multi-attention (MA) to extract features at different levels, and then we get within-block features $\mathbf{E}_F \in \mathbb{R}^{N \times D}$. Then, we transform  $\mathbf{E}_F$, $\mathbf{E}_A$ and $\mathbf{E}_{\text{enc}}$ to within-block representations $\mathbf{E}_F^{Q_2}$,  $\mathbf{E}_A^{Q_2}$, $\mathbf{E}_{\text{enc}}^{K_2}$ and $\mathbf{E}_{\text{enc}}^{V_2}$ via MLP layer. Next, we obtain the queries, keys, and values before the second LN \& MA layer by the following function:
\begin{equation}
    \mathbf{Q}_2 = \mathbf{\beta}_3 \times \mathbf{E}_F^{Q_2} + \left ( 1 - \mathbf{\beta}_3 \right ) \times \mathbf{E}_A^{Q_2} \nonumber
\end{equation}
\begin{equation}
    \mathbf{K}_2 = \mathbf{E}_\text{enc}^{K_2} + \mathbf{PE}
\end{equation}
\begin{equation}
   \mathbf{V}_2 = \mathbf{E}_\text{enc}^{V_2} \nonumber
\end{equation}
where $\mathbf{PE}$ is sinusoidal positional embedding and $\mathbf{\beta}_3$ is  learnable parameter. Then, the output of the second LN \& MA layer is passed through the feed-forward network (FFN) to generate the next decode embedding. Finally, $\mathbf{E}_\text{dec}$ is obtained and sent to the output layer to predict target-speaker voice activities. The output layer is composed of a linear layer plus a sigmoid activation function, which can determine the length of decoding.

\vspace{-0.3cm}
\subsection{MA-MSE with Deep Interactive Module}
\label{IDM}
The memory-aware multi-speaker embedding (MA-MSE) module can  retrieve a clean and  discriminative multi-speaker embedding
from memory through a simple additive attention mechanism. In \cite{MAMSE}, the memory component is the core of the MA-MSE module, which is composed of many speaker embedding basis vectors extracted from additional datasets. Specifically, speaker embedding basis vectors can be obtained by clustering the speaker embeddings (e.g., i-vectors or x-vectors) and taking the cluster centers. Before feeding the features $\mathbf{F'}$ into MA-MSE module, we use a clustering-based approach to get the speaker activity $0/1$ mask $\mathbf{S} \in \mathbb{R}^{N \times T}$ on each frame. The features $\mathbf{F'}$ and mask $\mathbf{S}$ are multiplied to get the selected features $\mathbf{F}_S = \left [\mathbf{F}_S^1, \mathbf{F}_S^2, ..., \mathbf{F}_S^N \right ] ^T \in \mathbb{R}^{N \times D}$  for each speaker and a simple additive attention mechanism is used to choose the speaker embedding bases of the memory that are most similar to the current speech segment. 

After the CHiME-7 DASR Challenge, we found that if the structure of the MA-MSE module is not well-designed it may significantly impair performance in complex acoustic scenarios. In addition, a too simple mechanism also limits performance improvement. Based on this, we introduce the deep interactive module (DIM), which replaces the additive attention mechanism with a dot-product attention mechanism and deepens the number of interaction layers. This multi-scale feature fusion approach can better extract cleaner, more discriminating multi-speaker embedding from memory module. 

The DIM is composed of 3 DIM blocks, each consisting of two cross-attention structures along the feature dimension. Given that all speaker embedding basis vectors in memory module are represented by $\mathbf{M} \in \mathbb{R}^{K \times D_M}$, where $K$ is the number of vectors. In first DIM block, the input features of $n$-th speaker $\mathbf{F}_S^n$ and $\mathbf{M}$ are used and calculated by the following function: 
\begin{equation}
    \mathbf{H}_1^n = \text{Softmax}\left(\frac{\left(\mathbf{F}_S^n\mathbf{W}_1^{n,q}\right)\left({\mathbf{M}\mathbf{W}_1^{n,k}}\right)^T}{\sqrt{d_m}}\right) \mathbf{M}
\end{equation}
where $\mathbf{W}_1^{n,q} \in  \mathbb{R}^{D \times D}$ and $\mathbf{W}_1^{n,k} \in  \mathbb{R}^{D_M \times D}$ are the learnable weights, the scaling $\sqrt{d_m}$ is for numerical stability. Then, the output of DIM block is calculated by:
\begin{equation}
    \mathbf{H}_2^n = \text{Softmax}\left(\frac{\left(\mathbf{F}_S^n\mathbf{W}_2^{n,q}\right)\left({\mathbf{H}_1^n\mathbf{W}_2^{n,k}}\right)^T}{\sqrt{d_m}}\right) \mathbf{H}_1^n
\end{equation}
where $\mathbf{W}_2^{n,q} \in  \mathbb{R}^{D \times D}$ and $\mathbf{W}_2^{n,k} \in  \mathbb{R}^{D_M \times D}$ are the learnable weights. After that, $\mathbf{H}_2^n$ and $\mathbf{F}_S^n$ are passed to the next DIM block. Finally, the MA-MSE embedding $\mathbf{E}_M$ is obtained.  $\mathbf{E}_{M}$, serving as essential supplementary speaker information, will be concatenated with the current speaker's i-vector to generate aggregate embedding $\mathbf{E}_{A}$. 
 \vspace{-0.3cm}
\subsection{Loss Function}
With the input acoustic features set $\mathbf{X}$ and speaker mask matrix, NSD-MS2S predicts speech/silence probabilities for each of the $N$ speakers with $\mathbf{\hat{Y}} = \left [ \hat{y}_\text{nt} \right ]_{N \times T}$. We adopt the binary cross-entropy (BCE) loss of multiple speakers as the learning objective:
\begin{equation}\label{eq:L1Loss}
\begin{aligned}
  \mathcal{L}=-\frac{1}{T}\sum_{t=1}^{T}\sum_{n=1}^{N}[y_{\text{nt}}\log(\hat{y}_{nt})+(1-y_{\text{nt}})\log(1-\hat{y}_{\text{nt}})]
\end{aligned}
\end{equation}

\section{Experiment}
\label{sec:Experiment}
\subsection{CHiME-7 DASR Challenge}
In CHiME-7 Challenge \cite{cornell2023chime},  Distant Automatic Speech Recognition (DASR) with Multiple Devices in Diverse Scenarios is the main track participants must always submit. The entire training data originates from
two parts, one is the CHiME-7 DASR “Official” data (e.g.,  CHiME-6, DiPCo, and Mixer 6 Speech), and
the other is the external data allowed under official rules. More details on how we organized and simulated the training data can be found in our technical report \cite{wang2023ustc}.

\vspace{-0.2cm}
\subsection{Implementation Details}
In our experiments, top-6 audio channels \cite{cornell2023chime} are selected to perform VAD using a
baseline VAD model fine-tuned by CHiME-6 and Mixer 6 data, we use ECAPA-TDNN \cite{desplanques2020ecapa} to extract x-vectors and spectral clustering (SC) as our diatization initialization system. 

For our NSD-MS2S system,  the input is the 40-dim FBANKs features. All encoder-decoder modules have 6 blocks sharing the same settings: the size of $D$ is 512, attention with 8 heads, and 1024-dim feed-forward layers with a dropout rate of 0.1. 
Conformer block is designed identically to \cite{gulati2020conformer}. The feature fusion coefficient $\beta$ is all initiated to 0.5. We use two MA-MSE modules whose memory modules consist of 256-dimensional x-vectors and 100-dimensional i-vectors, respectively. All speaker embeddings in the memory module are extracted from the VoxCeleb1 and VoxCeleb2 datasets. It is worth noting that the model we used for the CHiME-7 competition is without the DIM module. The decoding duration of the output layer is set to 8s ($T$=800) and mixup \cite{mixup} is used in training. We use Adam with a learning rate of $1e-4$ to optimize the entire model for 6 epochs. We found that even though there is a single-model performance degradation during training, there is still a gain when fusing models from different epochs.

\subsection{Results and Analysis}
\begin{table}
\centering
\setlength{\abovecaptionskip}{0pt}%
\setlength{\belowcaptionskip}{10pt}%
\caption{Performance comparison  of different methods on CHiME-7 DEV and EVAL set (collar = 0.25 s).}
\label{tab:single_model_results}
\begin{adjustbox}{max width=\linewidth}
\begin{tabular}{lccccccccc}
\toprule
\multicolumn{1}{c}{\multirow{2}{*}{Method}} & \multicolumn{1}{c}{\multirow{2}{*}{Set}} & \multicolumn{2}{c}{CHiME-6} & \multicolumn{2}{c}{DiPCo} & \multicolumn{2}{c}{Mixer 6} & \multicolumn{2}{c}{Macro}\\
\cmidrule(r{4pt}){3-4}
\cmidrule(r{4pt}){5-6}
\cmidrule(r{4pt}){7-8}
\cmidrule(r{4pt}){9-10}
\multicolumn{1}{c}{} & \multicolumn{1}{c}{} & \multicolumn{1}{c}{\textbf{DER}} & \multicolumn{1}{c}{\textbf{JER}}  & \multicolumn{1}{c}{\textbf{DER}} & \multicolumn{1}{c}{\textbf{JER}} & \multicolumn{1}{c}{\textbf{DER}} & \multicolumn{1}{c}{\textbf{JER}} & \multicolumn{1}{c}{\textbf{DER}} & \multicolumn{1}{c}{\textbf{JER}} \\
\midrule
\multicolumn{1}{l}{\multirow{2}{*}{x-vectors + SC}} & DEV & 40.32 & 42.31 & 24.47 & 28.97 & 15.8 & 23.07 & 26.86 & 31.45\\
\multicolumn{1}{c}{} & EVAL  & 36.32 & 43.39 & 25.18 & 35.08 & 9.53 & 12.08 & 23.67 & 30.18  \\
\cmidrule(r{4pt}){2-10}
\multicolumn{1}{l}{\multirow{2}{*}{+ NSD-MA-MSE}} & DEV  & 32.27 & 34.76 & 21.04 & 24.01 & 9.28 & 12.94 & 20.86 & 23.90 \\
\multicolumn{1}{c}{} & EVAL  & 32.09 & 37.61 & 22.78 & 31.34 & 6.21 & 7.12 & 20.36 & 25.35  \\
\cmidrule(r{4pt}){2-10}
\multicolumn{1}{l}{\multirow{2}{*}{+ NSD-MS2S}} & DEV  & 29.93 & 33.92 & 18.22 & 22.36 & 9.85 & 13.08 & 19.33 & 23.12  \\
\multicolumn{1}{c}{} & EVAL  & 30.50 & 36.01 & 21.64 & 29.83 & 5.5 & 6.3 & 19.21 & 24.04 \\
\cmidrule(r{4pt}){2-10}
\multicolumn{1}{c}{\multirow{2}{*}{+ DIM}} & DEV  & $\boldsymbol{28.36}$ & $\boldsymbol{31.49}$ & $\boldsymbol{17.06}$ & $\boldsymbol{18.54}$ & $\boldsymbol{7.27}$ & $\boldsymbol{9.56}$ & $\boldsymbol{17.56}$ & $\boldsymbol{19.86}$ \\
\multicolumn{1}{c}{} & EVAL  & $\boldsymbol{29.45}$ & $\boldsymbol{33.84}$ & $\boldsymbol{19.31}$ & $\boldsymbol{26.63}$ & $\boldsymbol{5.01}$ & $\boldsymbol{5.54}$ & $\boldsymbol{17.92}$ & $\boldsymbol{22.00}$  \\
\bottomrule
\end{tabular}
\end{adjustbox}
\end{table}

\begin{table}[b]
\centering
\setlength{\abovecaptionskip}{0pt}%
\setlength{\belowcaptionskip}{10pt}%
\caption{Computational resource consumption with NSD-MA-MSE and NSD-MS2S.}
\label{tab:model_infer_cmp}
\begin{adjustbox}{max width=\linewidth}
\begin{tabular}{lccc}
\toprule
\multicolumn{1}{c}{\multirow{1}{*}{Method}} & \multicolumn{1}{c}{\multirow{1}{*}{Parameters (M) $\downarrow$}} & \multicolumn{1}{c}{\multirow{1}{*}{GPU Memory (MB) $\downarrow$}}  & \multicolumn{1}{c}{\multirow{1}{*}{Inference Time $\downarrow$}} \\
\midrule
\multicolumn{1}{l}{\multirow{1}{*}{NSD-MA-MSE}} &\multicolumn{1}{c}{\multirow{1}{*}{50.19}} & \multicolumn{1}{c}{\multirow{1}{*}{5803}}  & \multicolumn{1}{c}{\multirow{1}{*}{1.0}} \\
\midrule
\multicolumn{1}{l}{\multirow{1}{*}{NSD-MS2S}} &\multicolumn{1}{c}{\multirow{1}{*}{57.53}} & \multicolumn{1}{c}{\multirow{1}{*}{3019}}  & \multicolumn{1}{c}{\multirow{1}{*}{0.47}} \\
\multicolumn{1}{c}{\multirow{1}{*}{+DIM}} &\multicolumn{1}{c}{\multirow{1}{*}{59.07}} & \multicolumn{1}{c}{\multirow{1}{*}{3050}}  & \multicolumn{1}{c}{\multirow{1}{*}{0.48}} \\
\bottomrule
\end{tabular}
\end{adjustbox}
\end{table}
Since Dover-lap \cite{raj2021dover} works slowly when the number of channels is large, we averaged the posterior probabilities of all the channels output by NSD-MS2S as the final output. Our diarization system is actually a multi-step iterative system in the CHiME-7 DASR Challenge \cite{wang2023ustc}, but for a fair comparison, we present the results of different single model systems at the first iteration in the Table \ref{tab:single_model_results}. Compared to NSD-MA-MSE , NSD-MS2S makes the macro DER drop relatively by 5.6$\%$ on EVAL set. Furthermore, DIM enhances the performance of the NSD-MS2S, resulting in a reduction of the macro DER from 19.21$\%$ to 17.92$\%$ on EVAL set.


In Table \ref{tab:model_infer_cmp}, we have analyzed the computational efficiency and overhead of the different methods under the same setting: batch is 16, one 12G Tesla V100 on the same machine. The inference time takes the average time to complete three runs of the CHiME-6 DEV set and is benchmarked using NSD-MA-MSE (with 1.0 as the reference). It can be seen that although the number of parameters has increased in NSD-MS2S, the GPU memory footprint and inference speed are superior to NSD-MA-MSE.

Table \ref{tab:fusion_models_results} shows the results of model fusion for 6 different epochs. The single model results of NSD-MS2S with DIM  are even better than the fusion models we used in the CHiME-7 DASR Challenge at first iteration (macro DER on EVAL set, 17.92 vs 18.32). However, the fusion models results of NSD-MS2S with DIM at first iteration are still worse than the final iteration we submitted to the CHiME-7 challenge (macro DER on EVAL set, 17.43 vs 15.87). On one hand, this demonstrates the effectiveness of our iterative strategy \cite{wang2023ustc} in enhancing the performance of NSD-S2S, with notable improvements across CHiME-6 set and DiPCo set, except for small performance degradation on the Mixer 6 set. On the other hand, it suggests the potential for surpassing our CHiME-7 champion system.


\begin{table}
\centering
\setlength{\abovecaptionskip}{0pt}%
\setlength{\belowcaptionskip}{0pt}%
\caption{Performance comparison of different fusion models on CHiME-7 DEV and EVAL set (collar = 0.25 s).}
\label{tab:fusion_models_results}
\begin{adjustbox}{max width=\linewidth}
\begin{threeparttable}
\begin{tabular}{lccccccccc}
\toprule
\multicolumn{1}{c}{\multirow{2}{*}{Method}} & \multicolumn{1}{c}{\multirow{2}{*}{Set}} & \multicolumn{2}{c}{CHiME-6} & \multicolumn{2}{c}{DiPCo} & \multicolumn{2}{c}{Mixer 6} & \multicolumn{2}{c}{Macro}\\
\cmidrule(r{4pt}){3-4}
\cmidrule(r{4pt}){5-6}
\cmidrule(r{4pt}){7-8}
\cmidrule(r{4pt}){9-10}
\multicolumn{1}{c}{} & \multicolumn{1}{c}{} & \multicolumn{1}{c}{\textbf{DER}} & \multicolumn{1}{c}{\textbf{JER}}  & \multicolumn{1}{c}{\textbf{DER}} & \multicolumn{1}{c}{\textbf{JER}} & \multicolumn{1}{c}{\textbf{DER}} & \multicolumn{1}{c}{\textbf{JER}} & \multicolumn{1}{c}{\textbf{DER}} & \multicolumn{1}{c}{\textbf{JER}} \\
\midrule
\cmidrule(r{4pt}){2-10}
\multicolumn{1}{c}{\multirow{2}{*}{Fusion${^\star}$}} & DEV  & 30.34 & 32.34 & 18.54 & 20.20 & 7.99 & 10.76 & 18.95 & 21.1 \\
\multicolumn{1}{c}{} & EVAL  & 29.39 & 34.13 & 20.58 & 28.67 & 5.01 & 5.57 & 18.32 & 22.79  \\
\cmidrule(r{4pt}){2-10}
\multicolumn{1}{c}{\multirow{2}{*}{Fusion${^\dagger}$}} & DEV  & 26.78 & 29.45 & 16.13 & 17.38 & $\boldsymbol{7.22}$ & $\boldsymbol{9.45}$ & 16.71 & 18.76  \\
\multicolumn{1}{c}{} & EVAL  & 28.51 & 32.63 & 18.83 & 25.72 & $\boldsymbol{4.95}$ & $\boldsymbol{5.45}$ & 17.43 & 21.26 \\
\cmidrule(r{4pt}){2-10}
\multicolumn{1}{c}{\multirow{2}{*}{Fusion${^\P}$}} & DEV  & $\boldsymbol{25.81}$ & $\boldsymbol{27.64}$ & $\boldsymbol{15}$ & $\boldsymbol{15.92}$ & 8.96 & 12.27 & $\boldsymbol{16.59}$ & $\boldsymbol{18.61}$ \\
\multicolumn{1}{c}{} & EVAL  & $\boldsymbol{25.11}$ & $\boldsymbol{28.86}$ & $\boldsymbol{16.36}$ & $\boldsymbol{22.06}$ & 6.14 & 6.81 & $\boldsymbol{15.87}$ & $\boldsymbol{19.25}$ \\
\bottomrule
\end{tabular}
\begin{tablenotes}
       \Large 
       \item ${^\star}$ and ${^\dagger}$ represent the results for NSD-MS2S and NSD-MS2S (+DIM) at the first iteration, respectively.
       \item ${^\P}$ stands for our NSD-MS2S system result submitted to  the main track of CHiME-7 DASR Challenge after final iteration. 
\end{tablenotes}
\end{threeparttable}
\end{adjustbox}
\end{table}

\section{Conclusions}
\label{sec:Conclusions}
We presented a novel approach called NSD-MS2S for the diarization of multi-speaker conversations, which provided state-of-the-art results in the CHiME-7 DASR Challenge. Compared to NSD-MA-MSE, NSD-MS2S not only increases the speed and reduces GPU memory consumption, but also improves performance. In the future, we will explore the directions of lighter structure, faster inference, and less resource consumption to facilitate the diarization system further toward practicality.

\small\bibliographystyle{IEEEbib}
\bibliography{mybib}
\end{document}